# Tandem Photovoltaics from 2D Transition Metal Dichalcogenides on Silicon

Zekun Hu[1], Sudong Wang[1], Jason Lynch[1], Deep Jariwala[1,*]

[1]Department of Electrical and Systems Engineering, University of Pennsylvania,

Philadelphia, PA, USA

**Keywords:** Superlattice, 2D semiocnductors, Excitons, solar cells, transition metal dichalcogenides.

## Abstract

The demand for high-efficiency photovoltaic systems necessitates innovations that transcend the efficiency limitations of single-junction solar cells. This study investigates a tandem photovoltaic architecture comprising a top-cell with a transition metal dichalcogenide (TMDC) superlattice absorber and a bottom-cell of crystalline silicon (c-Si), focusing on optimizing the light absorption and electrical performance of the combined structure. Through the transfer matrix method and electrical simulations, we optimized the geometry of the superlattice, determining that a six-layer $MoSe_2$ configuration with a 40 nm $SiO_2$ antireflective layer maximizes photon absorption while mitigating additional weight and preserving the cell's structural integrity. The results show that the optimized TMDC superlattice significantly improves the PCE of the tandem design to 30.94%, an increase of 7.66% over the original single-junction c-Si solar cell's efficiency. This advancement illustrates the potential of TMDC materials in next-generation solar cells and presents a promising avenue for the development of highly efficient, tandem photovoltaic systems vis van der Waals integration of the to-cell on Si.

## Introduction

The photovoltaic (PV) sector has experienced considerable expansion in recent years as a strategic response to mitigate carbon emissions (1). The industry has been predominantly spearheaded by crystalline silicon technology primarily due to its progressive cost reduction and efficiency enhancements (2). According to the Shockley-Queisser (SQ) limit (3), crystalline silicon (c-Si) is close to the ideal semiconductor for PVs with its band gap of 1.12 eV. In recent years experimental cell efficiency of up to 26.7% (4) has been demonstrated for c-Si cells which is slightly less than the SQ limit of 29.4% for c-Si. To further enhance performance, the industry is exploring various innovative technologies, including light-trapping and anti-reflection coatings (ARC) (5), passivation techniques (2), heterojunctions (6,7), and tandem structures (8). Notably, the integration of c-Si with other absorbing materials to form tandem solar cells presents a compelling avenue for enhanced performance, offering scalability and the potential to surpass the efficiency limitations of conventional c-Si solar cells. This approach signifies a strategic pivot in solar technology, aiming to combine the established strengths of c-Si with novel materials to achieve higher efficiencies and broader applicability.

In optimized, single-junction PVs, the largest loss mechanism is thermalization as high-energy photocarriers dissipate energy to relax to the band edge before being collected as electrical energy (9). Around 40% of all solar energy is lost through thermalization in silicon PVs. Tandem solar cells reduce thermalization loss by combining semiconductors with different band gaps in a stacked configuration such that high-energy photocarriers can relax to the band edge of the higher-band gap semiconductor without dissipating as much energy. With a large enough number of semiconductors, the theoretical power conversion efficiency (PCE) increases from 32.5% to 88% (10). Recent advances in tandem devices of perovskites and Si have produced cells with efficiencies of more than 31% (11,12). These developments not only demonstrate the potential of c-Si based tandem solar cells in maximizing energy harvest from the sun but also open pathways for cost-effective and sustainable solar energy solutions.

Transition metal dichalcogenides (TMDCs), specifically compounds like $MoS_2$, $WS_2$, $MoTe_2$, $MoSe_2$, and $WSe_2$, are gaining prominence as potential materials for thin-film PV applications (13-15). Their unique appeal lies in their transition from an indirect to a direct bandgap when reduced to monolayer thickness, facilitating high photoluminescence and radiative efficiency (16). This direct bandgap, ranging between 1 to 2.5 eV (17) includes the ideal band gap to pair with silicon in a tandem cell (1.7 eV (18)). Another distinctive feature of TMDC solar cells is the high absorption coefficient due to the dominance of excitonic physics (19), and their large refractive indices which enable ultra-compact photonic devices. The tunable electronic properties and ease of creating heterostructures with TMDCs, due to their van der Waals bonding, enable innovative tandem solar cell designs. In comparison to perovskites, which are known for their high power conversion efficiencies but suffer from stability issues (20), TMDCs offer the advantages of ultra-thin, lightweight structures, high-stability, and scalability, critical for applications in aerospace and wearable technology (21).

Building on the above discussed properties, we propose integrating a superlattice structure of monolayer TMDCs with c-Si to realize tandem solar cells. In our previous research (19), we extensively explored the viability of using TMDC superlattice structures, particularly focusing on $MoS_2$, for PV applications. Our prior investigation delved into key aspects such as the excitonic mechanisms, the influence of exciton binding energy, exciton lifetimes, exciton diffusivity, and free carrier mobility. Building upon this foundational work, the current study expands our understanding by considering the discrepancies caused by binding energy between the optical and electrical bandgaps and the performance in a tandem structure with c-Si bottom cell. Additionally, we have broadened our scope to include a comparison of different TMDC materials and integrating more experimentally-reported values to ensure a more accurate and realistic assessment. We present a model of a tandem solar cell, comprising a top-cell of a TMDC superlattice lateral homojunction with ARC and a bottom-cell of c-Si homojunction. Our simulations reveal that the TMDC superlattice top-cell, significantly contributes to enhancing the overall PCE. Specifically, the PCE contributions are 12.43% from the top-cell and 18.51% from the bottom-cell, culminating in a total PCE of 30.94%. This represents a substantial improvement compared to a simulated single-junction c-Si solar cell, which achieves a PCE of 23.28%. Our findings underscore the potential of TMDC superlattice structures as top-cell layers in boosting the efficiency of conventional c-Si solar cells, opening new avenues for high-efficiency, lightweight solar energy solutions.

## Structure and Simulation method

The proposed tandem solar cell is shown in Figure 1a and combines a top cell of superlattice of alternating layers of monolayer TMDCs ($MoS_2$, $MoSe_2$, $WS_2$, $WSe_2$, and $MoTe_2$) and an insulating material ($Al_2O_3$ and h-BN) with a c-Si bottom cell and an insulating intermediate layer that is used to enhance the absorption in the superlattice. The proposed superlattice structures have been experimentally demonstrated by our group at $cm^2$ scale in prior work (22). The top cell is configured in a horizontal p-i-n junction. The total length of the simulated top cell is 1 µm, with either the p- or n-region comprising 1% of this length, situated adjacent to the cell boundaries. Uniform doping of $10^{19}$ $cm^{-3}$ is applied on each side to facilitate the junction formation with the cathode (Ag) and anode (Au) at either end. Separating the top TMDC cells from the bottom c-Si cell is a spacer layer ($SiO_2$ and $Al_2O_3$) whose thickness has been optimized to maximize absorptance. The bottom cell features a vertically oriented n-p silicon structure with a rear contact of Al and a front contact of Au. The bottom cell is further enhanced by a 75 nm nitride layer on the top surface with n-type front doping, aimed at augmenting its overall performance.

The transfer matrix method (TMM) was employed using Python to accurately model the optical properties of the PV system (23). The TMM uses refractive index values from literature (24-26) to simulate the absorptance spectra of the one-dimensional system. The photocarrier generation rate is then calculated by multiplying the absorptance spectrum and the AM 1.5G solar spectrum (27). This approach provides a realistic estimation of the solar cell's optical performance under typical solar conditions. Additionally, for instances of non-normal light incidence, the absorption spectra for transverse electric (TE) and transverse magnetic (TM) polarizations were averaged to model unpolarized solar light.

The electronic performance of the tandem solar cell was simulated using Sentaurus TCAD which is commonly used to model PVs. The simulation encompasses the entire structure of the top cell, including the active, monolayer TMDCs, the insulating layers, and the electrodes. Notably, the thick $SiO_2$ spacer layer of the bottom cell was excluded from the simulation due to its electrical insulating properties. The simulation model integrated the Poisson equation, singlet exciton equation, and continuity equation in a finite element model, thereby fully accounting for the excitonic behaviors characteristic of TMDC materials. This comprehensive approach allowed for the simulation of exciton generation, diffusion, and recombination processes, as well as carrier transport and recombination within the cell. The parameters for each absorbing material were finely tuned based on optimizations from prior research(19). In this model, electron and hole densities were calculated using the respective quasi-Fermi potentials. The band gaps were determined based on reported empirical values, and details are provided in supporting information (SI) Section 4. The model was particularly designed to accommodate discontinuous interfaces, characteristic of a superlattice structure. For the exciton dissociation process, 10 interfaces – including those between the p-/i-region, i-/n-region, and within the i-region – were specifically modeled. Detailed simulation parameters and additional information are provided in SI Section 1.

The bottom Si component of the tandem solar cell was simulated independently using TCAD Sentaurus (28,29). For this part of the simulation, a three-dimensional c-Si solar cell with a rear contact design was modeled. The simulation process involved the computation of an optical generation profile, current-voltage data, and key PV parameters such as open-circuit voltage ($V_{OC}$), short-circuit current ($J_{SC}$), fill factor (FF), and PCE. In this model, particular attention was paid to the differing reflectance between the rear contacts and the uncovered regions of the cell. Additionally, it was assumed that photon generation beneath the front contact is negligible. The model also accounted for the impact of doping levels on the mobility of free carriers within the silicon substrate. The substrate itself is lightly doped with phosphorus ($10^{15}$ $cm^{-3}$), while the front surface features a heavily doped n+ layer (with a peak concentration of $3\times10^{19}$ $cm^{-3}$ at the boundary of the top contact). Near the rear contact, p+ doping is also set at $3\times10^{19}$ $cm^{-3}$. A nitride antireflective layer covers the surface, enhancing the cell's overall efficiency. For comprehensive insights into the simulation parameters and additional details, reference is made to the SI Section 2.

## Results and Discussion
### Absorption Characteristics
In the tandem TMDC/c-Si solar cell configuration, shown in Figure 1a-c, the top TMDC superlattice and bottom c-Si cells function as electrically independent units but are optically interconnected. The structure of the top cell consists of repeating layers: each monolayer TMDC absorbing layer is 0.7 nm thick, interspersed with 3 nm of $Al_2O_3$ as an insulator which is sufficiently thick to electronically isolate photocarriers to a single layer (22). The choice of monolayer TMDC materials capitalizes on their direct bandgap property, which is advantageous over their bulk counterparts (30). For the designed 6-layer superlattice top cell, the cumulative thickness of the active layers is 4.2 nm. In contrast, the bottom c-Si solar cell maintains a substantial thickness of 150 µm. This combination of the ultra-thin top layer with the thicker bottom c-Si cell utilizes the high power density attribute of the top TMDC superlattice, a feature that has been previously reported (19). The combination of these design elements in the TMDC/c-Si tandem solar cell aims to harness the unique optical properties of TMDCs while leveraging the established efficiency and robustness of c-Si technology.

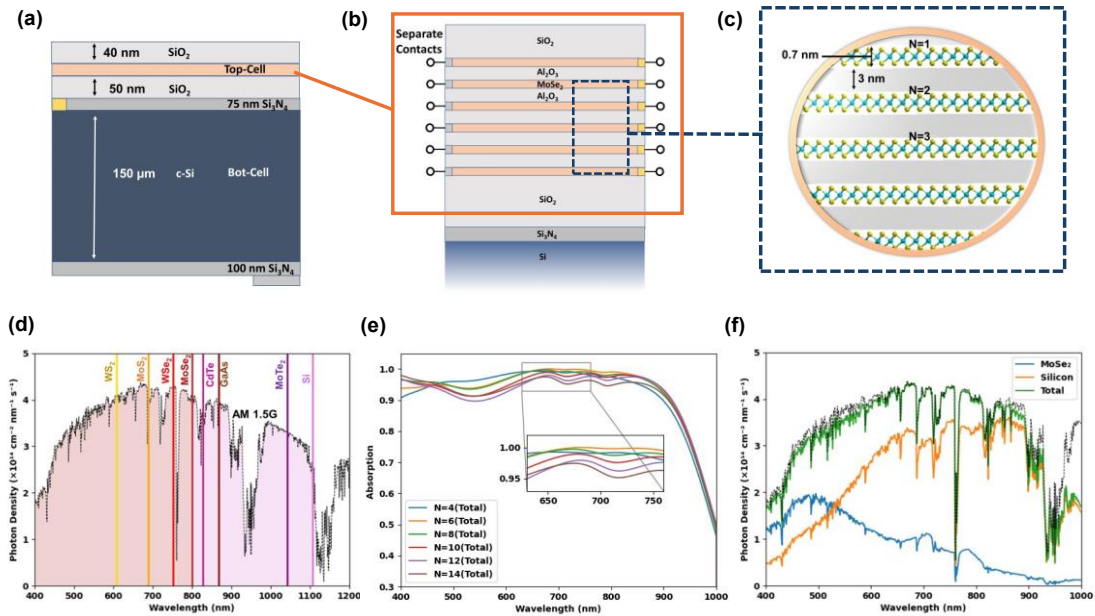

Figure 1. Comprehensive characterization of a TMDC superlattice and c-Si hybrid solar cell. (a) Schematic of TMDC superlattice/c-Si solar cell. The system comprises a top cell composed of periodically horizontal monolayer solar cell and a c-Si rear contact bottom cell. (b) A 6-layer $MoSe_2$ top-cell featuring independent contact interfaces. (c) Detailed monolayer structure in the superlattice (N represents the number of layers). (d) Solar spectrum of the sun (AM 1.5G) and band gaps of common solar cell absorbing materials. (e) Total absorption spectrum of the solar cell based on $MoSe_2$ top layer. (f) Photon absorption density in a 6-layer $MoSe_2$ superlattice and c-Si solar cell in each cell and total relative to the AM 1.5G solar spectrum.

In accordance with the SQ model, the band gaps of semiconductors in a tandem cell are critical in determining the theoretical maximum PCE of solar cells. Considering the range of bandgaps for common industrial solar cell materials— Si (1.12 eV), InP (1.35 eV (31)), and CdTe (1.5 eV (32))—the integration of monolayer TMDC materials with bandgaps spanning from 1.19 eV to 2.04 eV offers a pathway to fabricating efficient tandem solar cells (33), since this range encompasses the ideal band gaps to pair with these well-developed solar cell materials (18). C-Si is very close to the lower end of band-gap values in this range. Further it is the most scaled up and dominant solar cell technology justifying our choice. Figure 1d illustrates the design principle of tandem cells. $MoSe_2$, with its higher bandgap, is positioned as the top-cell to absorb higher energy photons, while the c-Si bottom cell captures lower energy photons. This configuration is visually represented, where the red region indicates the ideal absorption range of $MoSe_2$, and the pink region corresponds that of c-Si. This strategic cell configuration is engineered to mitigate thermalization losses, thereby optimizing the conversion of solar to electrical energy.

As discussed in more detail below, a $MoSe_2$ superlattice showed the best performance upon integration with c-Si. Upon selecting $MoSe_2$ and c-Si for the tandem solar cell, it is crucial to assess the actual absorption performance within the framework of a superlattice structure, particularly when the total active top layer is only several nanometers thick. Figure 1e presents the total absorption profile of the tandem cell across a tested spectrum (λ = 400 to 1,000 nm) for various number of unit cells in the superlattice (N = 4 to 14). For wavelengths shorter than 850 nm, the total absorption of the tandem cell is observed to have near-unity absorption. For wavelengths in 650-750 nm, where the solar intensity is relatively high, a superlattice with 6 layers shows best performance in absorption. Based on these findings, the superlattice was optimized to 6 layers for this study, thereby harmonizing the need for a thin active top layer with the imperative of efficient absorption across the pertinent solar spectrum.

The integration of the AM 1.5G solar spectrum with tailored absorption spectra of the chosen materials enables a comparative analysis of the absorption profiles for the tandem solar cell's top and bottom cells, as depicted in Figure 1f. Notably, the top cell, which utilizes a superlattice structure, exhibits superior photon absorption for wavelengths below 540 nm when compared to the bottom c-Si cell. However, within the spectral range of 540-800 nm—where the top cell, theoretically designed with a 1.55 eV bandgap, should exhibit higher photon absorption—the six-layer $MoSe_2$ superlattice shows a suboptimal absorption rate. This less-than-ideal absorption efficiency in the top cell is likely a consequence of its ultra-thin superlattice configuration. For longer wavelengths, beyond 700 nm, where photon energy is lower, a deviation from ideal total absorption is observed. This is attributed to the inherent limitations of the planar c-Si solar cell design, which could potentially be mitigated by introducing a textured surface to increase light trapping and absorption (34).

## Geometry Optimization

The selection and dimensional optimization of the insulating layers are critical to maximizing the cell's performance. The insulator thickness between $MoSe_2$ layers, bottom insulating, and top insulating layers were optimized to maximize the absorbed photon energy, and the performance in terms of photon density is in SI (Figure 2 and SI Figure S1-3). Figure 2a investigates $Al_2O_3$ and h-BN as potential insulators to determine the optimal material for use between monolayers in the tandem solar cell structure. The purposes of the insulator are to act as a spacer that mitigates electrical coupling between the layers, to facilitate Coulomb engineering to tune exciton binding energy (35), and to enhance light absorption by exploiting the differences in refractive index (22). The analysis reveals that the absorption efficiency using an h-BN-based spacer declines sharply for thicknesses greater than 2 nm. In contrast, $Al_2O_3$ demonstrates a peak absorption efficiency within a thickness range of 3 to 4.5 nm. With the goal of maximizing total energy absorption while minimizing structural weight, a 3 nm thickness of $Al_2O_3$ was selected for the optimal insulator layer.

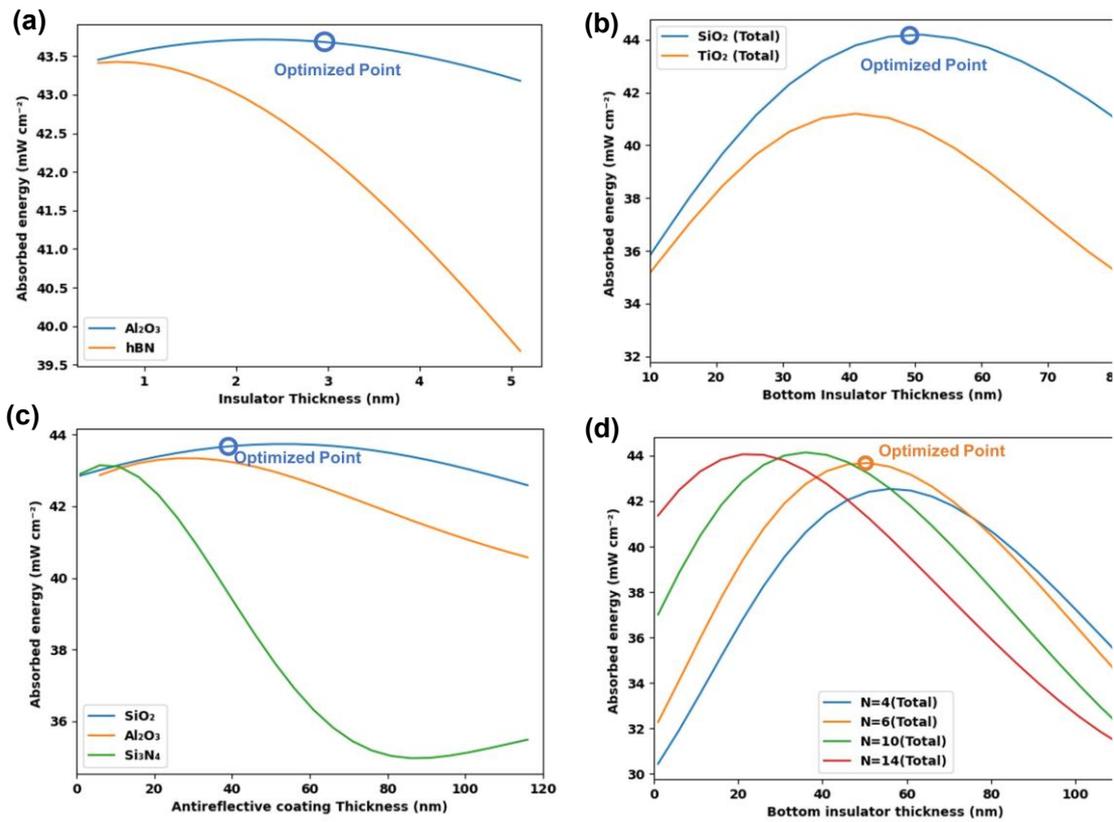

Figure 2. Optimizing Geometrical Parameters for Enhanced Light Absorption. (a) Comparative analysis of hBN and $Al_2O_3$ as insulators between $MoSe_2$ monolayers to maximize total photon absorption. (b) Evaluation of $SiO_2$ and $TiO_2$ insulators for optimal interfacing between the TMDC superlattice and the c-Si solar cell. (c) Optimization of top antireflective coating layers with various insulator materials. (d) Optimization of the insulator ($Si_3N_4$) between two cells and superlattice layers to augment absorption.

To design an optimal thick insulator layer between the top and bottom cells of the tandem solar cell, various materials were scrutinized. Initially, $TiO_2$ was considered due to its prevalent use as an antireflective layer in Si-based solar cells (36). However, upon evaluating the photon absorption densities of $TiO_2$ in conjunction with $SiO_2$, $SiO_2$ emerged as the superior insulator material, particularly at a 50 nm thickness, as depicted in Figure 2b. Although $TiO_2$ enhances the absorption in the Si bottom cell, its overall contribution is dampened by the suboptimal absorption in the TMDC top cell, as shown in SI Figure S3. Therefore, to achieve the best balance between top and bottom cell absorption, and hence maximize the tandem cell's overall efficiency, a 50 nm layer of $SiO_2$ was integrated as the insulator in the finalized cell architecture.

This imperfect photon absorption is primarily attributed to the reflective properties of $MoSe_2$ necessitating the incorporation of an ARC. This layer aims to mitigate photon loss by minimizing reflection at the $MoSe_2$ surface. The effect of an ARC can be further induced using nanostructures on top of the tandem PV (37). Figure 2c illustrates the impact of adding an ARC layer on the solar cell's absorption characteristics. It was observed that the overall cell performance improved with the introduction of the ARC layer, particularly for 40-60 nm of $SiO_2$. The orange and green lines denote the performance with $Al_2O_3$ and $Si_3N_4$ ARC layers, respectively. Considering the total absorbed energy, the desire to increase light capture in the $MoSe_2$, and the objective to add minimal weight to the top structure, a 40 nm $SiO_2$ ARC layer was selected as the optimal thickness. The addition of this minimal ARC layer resulted in a modest but notable increase of 0.8 mW cm$^{-2}$ in the total photon absorption for the tandem solar cell.

Optimizing the insulating layer that electrically separates the top TMDC superlattice from the bottom c-Si cell is a pivotal aspect of enhancing the tandem solar cell's performance. This insulating layer acts as an ARC for the bottom cell and facilitates resonance coupling, which varies with the thickness of the top cell superlattice. In this study, N in the top cell was varied from 4 to 14 to identify the configuration that maximizes photon energy capture. The results indicated that a 14-layer superlattice achieved the highest photon energy, with an absorption of 44.7 mW cm$^{-2}$. However, taking into account the complexities associated with fabricating such a multi-layered structure, a more practical design was adopted. The final configuration consists of a 6-layer $MoSe_2$ superlattice, coupled with a 50 nm thick $SiO_2$ insulating layer, achieving a near-optimal total absorption of 44.2 mW cm$^{-2}$. This design choice not only simplifies the fabrication process but also effectively balances high absorption efficiency with manufacturability. When translated into photon density, this configuration allows for $6.47 \times 10^{16}$ photons per cm$^2$ per second to be absorbed by the $MoSe_2$ superlattice, while the bottom c-Si solar cell captures $1.53 \times 10^{17}$ photons per cm$^2$ per second, illustrating the effectiveness of this optimized tandem solar cell architecture.

### Electrical Performance Evaluation

This section delves into the simulation of electrical characteristics of the cell, with a particular focus on the $MoSe_2$ layers. The top cell simulation incorporates the critical aspects of excitonic physics, including the generation, diffusion, recombination, and dissociation of excitons, to accurately model the electrical behavior of the TMDC material. To optimize the electrical output, each layer within the TMDC superlattice is configured with separate contacts to individually maximize electricity generation. As depicted in Figure 1b, this approach allows for precise control over the electrical properties of each layer. The performance of the superlattice is then simulated by evaluating each layer independently, based on its specific photon absorption density, to ensure an accurate representation of its contribution to the overall efficiency of the solar cell.

Figure 3a showcases the absorption spectrum of the top cell consisting of $MoSe_2$ layers. The monolayer $MoSe_2$, known for its direct bandgap transitions, exhibits strong excitonic absorption. It is observed that with fewer than six layers, the absorption at these peaks is not fully saturated, implying that the excitonic absorption capacity is not maximized. The comparison between the 6-layer and 4-layer configurations demonstrates a significant performance gap, suggesting that the addition of layers beyond the four-layer enhances the top cell's performance substantially. For N>6 designs, the strategic layer addition brings unnecessary complexity, albeit small enhancement in absorption efficiency.

In the electrical performance analysis of a 6-layer $MoSe_2$ superlattice, presented in Figure 3d, the focus is on the current-voltage (I-V) characteristics of each discrete layer within the superlattice structure. The I-V curve of the topmost layer is denoted as '1$^{st}$', with subsequent layers labeled sequentially down to the '6$^{th}$' or bottom layer. The I-V curves of the first four layers exhibit a high degree of similarity, including nearly identical short-circuit currents, indicative of uniform light absorption and conversion efficiency across these layers. However, a noticeable reduction in current is seen in the fifth and sixth layers due to the partial absorption of light by the preceding layers. The individual layers each contribute between 1.83-2.23% to the overall PCE, with the collective contribution of the six layers amounting to a PCE of 12.43%. The open-circuit voltage across the layers remains consistent, reflecting the uniform band gap of the MoSe$_2$ material throughout the superlattice.

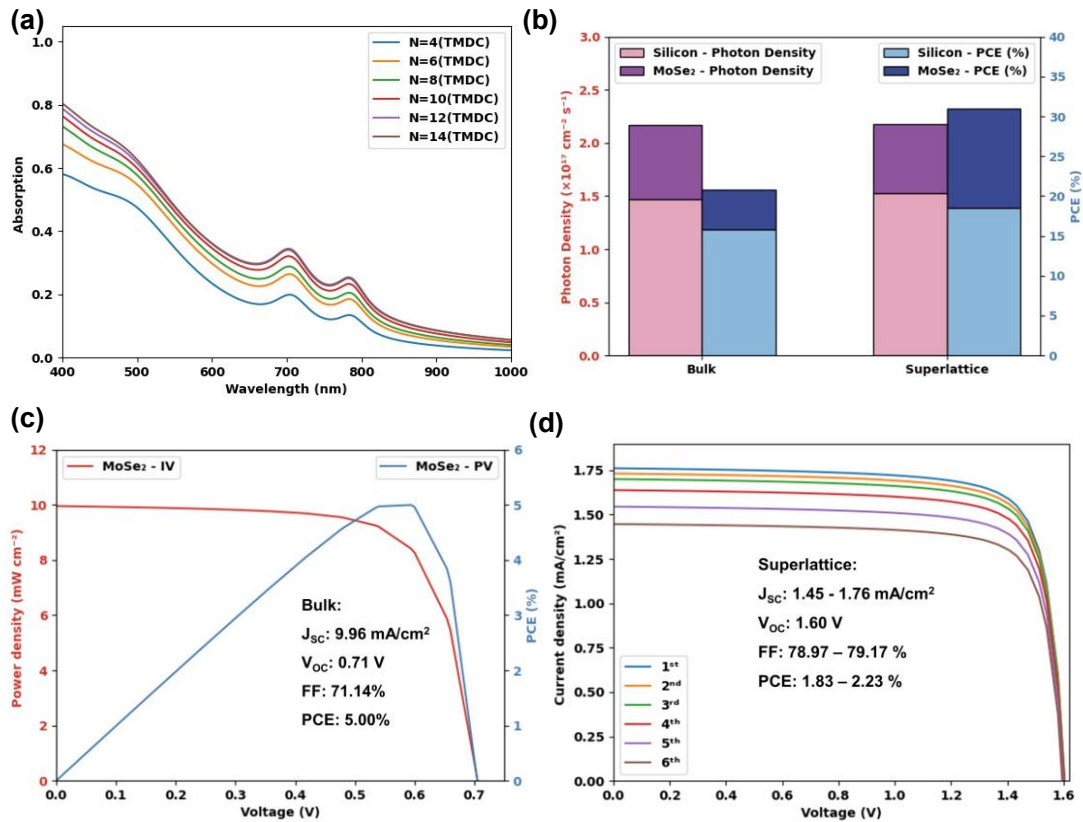

Figure 3. Optimization of the Superlattice Top Cell Structure. (a) Spectral absorption characteristics across varying numbers of TMDC layers in the top cell. (b) Photon absorption and PCE between superlattice and bulk-heterojunction top cells. (c) I-V and top-cell power-voltage performance profiles for the bulk top cell. (d) I-V performance profiles for individual superlattice layers.

The integration of a superlattice structure in tandem solar cells, while adding complexity, raises the question of its advantages over traditional bulk counterparts. Figure 3b addresses this by comparing the simulated performances of a 4.2 nm thick bulk MoSe$_2$ layer with a six-layer MoSe$_2$ superlattice. For bulk MoSe$_2$ when comparing to its monolayer form, the difference of optical constants, binding energy, and band gap have been considered. With the TMM and distinct optical constants for the bulk and two-dimensional forms, the photon density for the bulk structure is calculated at $6.97 \times 10^{16}$ cm$^{-2}$ s$^{-1}$, which is comparable to the superlattice photon density of $6.47 \times 10^{16}$ cm$^{-2}$ s$^{-1}$. The performance is verified through electrical simulations, which indicate PCE of 5.00% and 12.43% for the bulk and superlattice MoSe$_2$, respectively. Since the two systems have comparable photon density, but the superlattice has a larger PCE, we can attribute the improved performance to the improved electrical qualities of monolayer $MoSe_2$. This includes both the direct band gap of monolayer TMDCs over their indirect bulk properties, and the more ideal electrical band gap in monolayer $MoSe_2$ (1.55 eV) than the bulk value (1.1 eV) for integration into a tandem cell with Si(18). This observation of improved PV performance in monolayer TMDCs over bulk is also consistent with previous theoretical work on the efficiency of ultra-thin TMDC PVs when the photon density is constant between TMDC thicknesses (38). Figure 3c shows the IV curve and power-voltage curve of the bulk $MoSe_2$ with a thickness of 4.2 nm.

Figure 3d shows the I-V curves of the top cell. The high performance of the solar cell, characterized by a notably high Voc, can be attributed to the unique electrical properties of $MoSe_2$. Specifically, the effective bandgap for electrical applications in this context is calculated as the sum of $MoSe_2$'s optical bandgap (1.55 eV) and its exciton binding energy (0.57 eV). This ideal scenario assumes negligible contact resistance and considers the potential for tuning the binding energy. Additionally, the implementation of a short channel length of 1 µm, essential for maintaining the high open-circuit voltage (Voc), presents a trade-off with respect to scalability. This configuration, while beneficial for enhancing Voc, may limit the potential for larger-scale applications due to challenges associated with maintaining consistent performance over larger areas. The results show the capability of TMDC materials to achieve a Voc exceeding 1 V, which has been demonstrated in other experimental studies (39,40). Such high Voc values contribute significantly to the overall efficiency of the tandem PV devices, showcasing the potential of TMDCs in high-performance PV applications. In practical settings, factors such as material defects, longer electrode distances, and non-ideal contact connections are likely to result in a measured Voc that is lower than the theoretical maximum. These practical considerations must be accounted for in experimental designs to closely approximate and understand the performance metrics under typical operating conditions.

In Figure 4a, a detailed view of the rear-contact c-Si solar cell is presented, extracted from a larger array of solar cells. This configuration highlights the rear surface, which is equipped with point contacts designed to minimize the contact area while balancing a set of competing physical phenomena. The variance in optical reflectivity and surface recombination velocities at the interface of silicon with aluminum, and the passivated interface with silicon nitride, indicates that a reduced rear contact area could enhance the cell's efficiency. However, this reduction must be carefully managed to mitigate associated performance losses due to current crowding, increased contact resistance, and bulk recombination that are exacerbated as the contact area diminishes. Through a process of optimization, the rear contact area has been fine-tuned to 1250 µm², set within a cell of 175×500 µm². In parallel, the design accounts for the low transmission through the top contact, opting for a smaller dimension of 175×10 µm² to maximize light entry. The model incorporates a comprehensive set of parameters, which are detailed in the SI document Section 5, to ensure the optimized performance of the rear-contact c-Si solar cell within the tandem configuration.

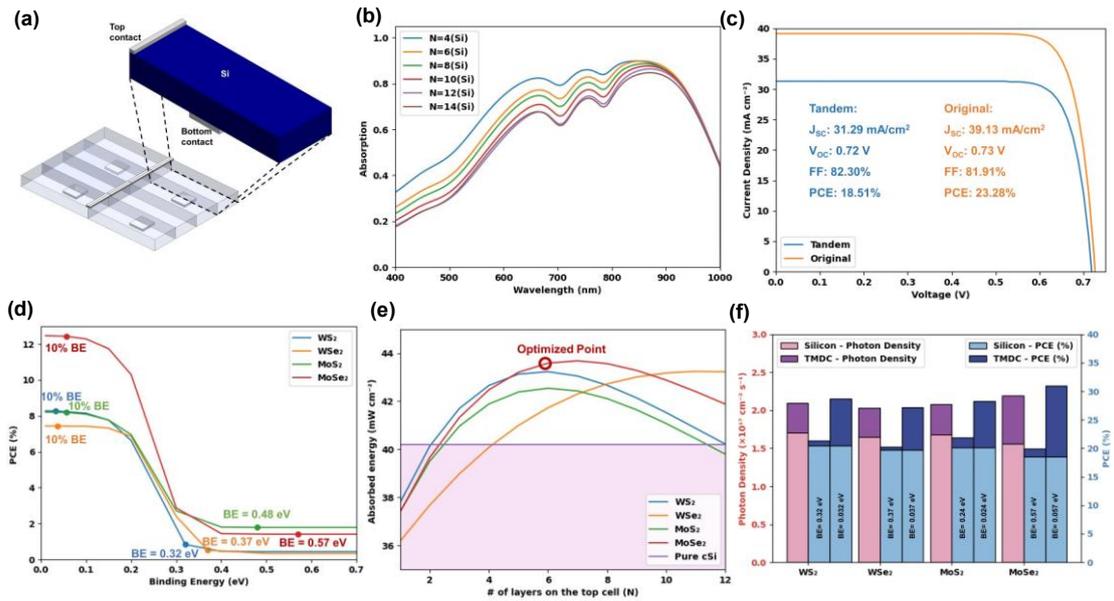

Figure 4. Bottom Cell Analysis and TMDC Material Comparison. (a) Schematic of the rear-contact c-Si solar cell configuration. (b) Absorption spectra variation of the Si bottom cell with different TMDC layer counts. (c) Comparative I-V characteristics of the original single-junction and the tandem solar cell configurations. (d) Top-cell PCE of various TMDC materials as a function of binding energies. Each TMDC's reported binding energy and a 10% adjusted value are labelled to highlight their impact on PCE. (e) Total absorbed energy density of various TMDC materials as a function of the number of superlattice layers, compared with a single-junction c-Si solar cell. (f) Correlation of photon absorption rates and PCE across various TMDC materials with distinct binding energy values.

Figure 4b depicts the absorption spectrum of the c-Si portion within a tandem solar cell structure, highlighting the layers of the $MoSe_2$ superlattice in response to different wavelengths of light. The I-V characteristics of the c-Si part in both the tandem configuration and the original single-junction solar cell are contrasted in Figure 4c. It is observed that the original single-junction cell exhibits a higher absorbed photon density in the Si layer, as detailed in Table 1, which leads to an elevated Jsc. This comparison verifies the impact of tandem cell architecture in photon absorption and electrical output.

The performance of various TMDC materials ($WS_2$, $MoS_2$, $WSe_2$, and $MoSe_2$) in a complete solar cell setup warrants comparative analysis due to their distinct electrical characteristics. As depicted in Figure 4d, the varying of exciton binding energy emerges as a pivotal factor in optimizing TMDC solar cell performance. This aspect of research is essential for guiding advancements within the TMDC solar cell community, pushing towards more competitive outcomes. Experimental values for the binding energy in free-standing monolayers ranging from 0.32 to 0.57 eV have been observed. However, the binding energy values are highly-sensitive to dielectric screening, and it has been found that binding energies can be reduced by ~90% upon encapsulation with $Al_2O_3$ ($\varepsilon_{static}$=9.8) on top and bottom of the monolayer due to dielectric screening of the fringing fields of the in-plane exciton (35). As such, each TMDC material reduced to 10% of its freestanding binding energy (41-43), and $MoSe_2$ demonstrates superior performance.

Figure 4e compares the energy absorption of superlattice tandem solar cells with that of traditional single-junction c-Si solar cells. The absorbed energy is calculated by multiplying the absorbed photon density by the absorber's bandgap. Results show that for top cells with N<4, the total absorbed energy falls below that of the pure c-Si cell. The optimal performance is observed at N=6 with $MoSe_2$, where it surpasses the c-Si solar cell. However, for N>8, a decreasing trend suggests that reflections from the top cell layers begin to detract from overall efficiency and causing a net reduction for most TMDCs.

As Figure 4f indicates, despite the similar total light absorption across all four materials, $MoSe_2$ exhibits a higher proportion of absorbed light being converted into electrical energy, represented in pink and purple, attributable to its relatively lower bandgap. In selecting the optimal TMDC for the tandem solar cell structure, $MoSe_2$ stands out due to suitable bandgap and electrical properties, which are critical for efficient tandem design. This choice is further justified in the electrical simulation, presented in blue, where the solar cell's performance is analyzed in the context of excitonic binding energy. Previous research has suggested that solar cell performance can be significantly influenced by the material's excitonic binding energy (19). For the simulation, the binding energy is considered at 10% of the typical exciton binding energy for a freestanding monolayer, reflecting adjustments through Coulomb engineering within the localized dielectric environment. $MoSe_2$, known for its suitable absorption bandgap and relatively large electrical bandgap when considering contributions of excitons, consequently exhibits the highest PCE among the TMDCs, for the tuned binding energy scenarios (other parameters in the SI Section 4). The deployment of a 6-layer $MoSe_2$ superlattice culminates in the performance metrics listed in Table 1. The tandem structure not only increases the total photon absorption but also raises the absorbed photon energy, thereby enhancing the overall PCE by 7.66% to a final value of 30.94%.

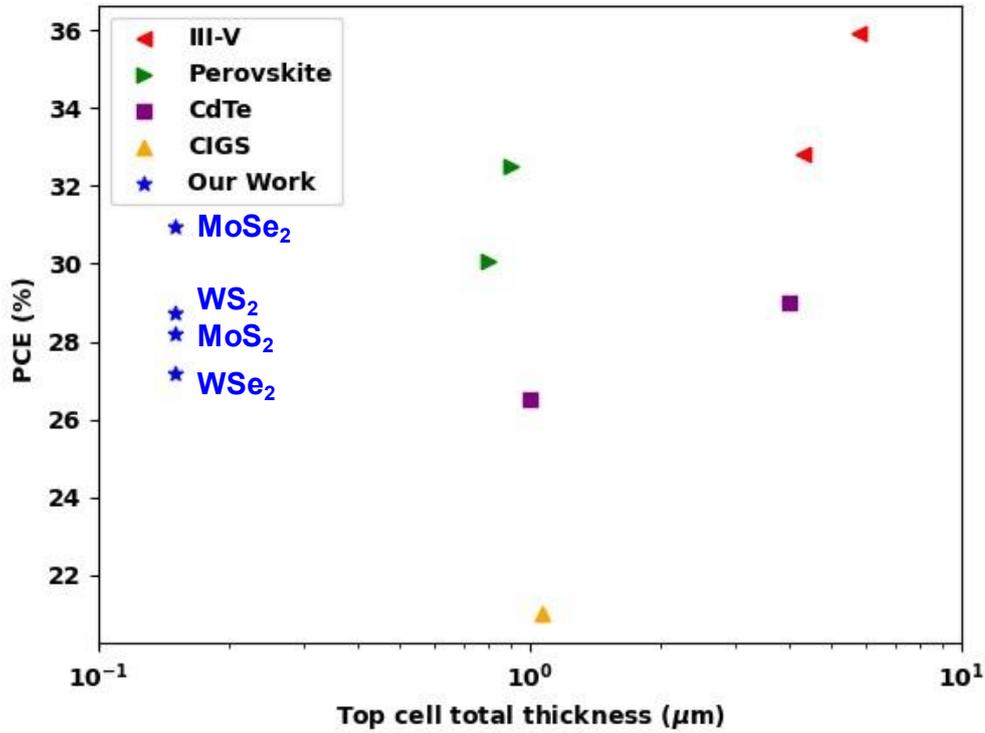

Figure 5. Efficiency chart of Si bottom cell-based tandem PV devices, comparing with III-V (44), Perovskite (45,46), CdTe (47), and Copper indium gallium selenide (CIGS) (48) top cells.

Figure 5 illustrates the competitive potential of our TMDC-based tandem solar cell designs with c-Si bottom cell against some of the most advanced c-Si bottom cell based tandem PV technologies in terms of PCE. Our work, marked distinctly with stars, stands out particularly in the landscape of III-V, perovskite, CdTe, and CIGS technologies, demonstrating competitive advancements in efficiency. Moreover, our approach maintains a total top cell thickness of only 150 nm, including the ARC, insulators, and spacers, which is much less than other tandem PVs because of the atomic thickness of the $MoSe_2$ active layers. This ultra-thin layering not only ensures high power density, but it also supports a lightweight design, making it an attractive option for enhancing commercial silicon solar cells. This slim profile is crucial for applications where weight and efficiency are paramount, offering substantial improvements over traditional PV technologies.

Table 1. Electrical Performance Metrics of $MoSe_2$ Superlattice and Silicon Parts in a Tandem Solar Cell Configuration.

|      | Superlattice layer | Jsc (mA cm$^{-2}$) | Voc (V) | FF (%) | PCE (%) |
| --- | --- | --- | --- | --- | --- |
| TMDC | 1st | 1.761 | 1.604 | 78.97 | 2.231 |
|      | 2nd | 1.731 | 1.604 | 78.99 | 2.193 |
| (MoSe$_2$) | 3rd | 1.700 | 1.603 | 79.02 | 2.153 |
|  | 4th | 1.638 | 1.601 | 79.06 | 2.073 |
|  | 5th | 1.545 | 1.598 | 79.13 | 1.954 |
|  | 6th | 1.446 | 1.596 | 79.19 | 1.828 |
| TMDC Sum |  | 9.821 |  |  | 12.43 |
| Si |  | 31.29 | 0.719 | 82.30 | 18.51 |
| Tandem Sum |  | 41.11 |  |  | 30.94 |
| Original Si |  | 39.13 | 0.73 | 81.91 | 23.28 |

## Conclusion

In conclusion, this study presents a comprehensive analysis of the PV performance of a tandem solar cell architecture employing a TMDC superlattice atop a c-Si bottom cell. Through meticulous simulations, we have demonstrated the effectiveness of MoSe$_2$ as a superior top-cell material due to its high free carrier mobility and lowest bandgap for monolayers among the four primary semiconducting TMDCs. The geometrical optimization of the superlattice, including the precise thickness of the insulating and antireflective layers, has been shown to be crucial for enhancing the overall absorption of the tandem cell. The optimized configuration utilizes a 6-layer MoSe$_2$ superlattice with a 40 nm Si$_2$O$_3$ antireflective layer, which collectively improves the photon absorption and contributes to a significant increase in PCE.

Electrical simulations have revealed that each layer of the MoSe$_2$ superlattice contributes effectively to the total PCE, with the 6-layer configuration achieving a PCE of 12.43%. The tandem cell structure, incorporating a rear-contact c-Si bottom cell optimized for size and contact area, not only preserves the high PCE inherent to silicon technology but also leverages the unique properties of the TMDC superlattice to achieve a combined PCE of 30.94%. This represents a substantial improvement over the original single-junction c-Si solar cell PCE of 23.28%. The insights from this study underscore the potential of TMDC materials in tandem solar cell applications and pave the way for future research aimed at realizing high-efficiency, lightweight solar cells for commercial and specialized applications.

## Acknowledgements

The authors acknowledge support from the Office of Naval Research (ONR) Young Investigator Award (YIP) (N00014-23-1-203) Metamaterials Program. D.J. also acknowledges support from the Alfred P. Sloan Foundation's Sloan Fellowship in Chemistry.

## Content of the supporting information

The Supporting Information accompanying our manuscript provides an exhaustive breakdown of the simulation models and optimization parameters used in our study.

Section 1 delves into the detailed modeling of the top-cell, featuring a TMDC superlattice, while Section 2 focuses on the bottom-cell, comprised of crystalline silicon. We include a series of supporting figures (Figure S1-S4) that detail the geometry optimization processes for various components such as the superlattice layers, antireflective layer, insulators, bottom-cell thickness, and top contact area, all centered around $WS_2$. Additionally, the precise simulation parameters are thoroughly documented across Tables S1-S4, offering a comprehensive reference for replication and further study. Figure S5 shows the optimization of free carriers, doping concentration, and length of p-/n- regions. Figure S6 presents the results from the simulated bottom-cell, showcasing the effectiveness of our modeling approach.

# Reference

(1) Asim, N.; Sopian, K.; Ahmadi, S.; Saeedfar, K.; Alghoul, M. A.; Saadatian, O.; Zaidi, S. H. A review on the role of materials science in solar cells. *Renewable and Sustainable Energy Reviews* **2012**, *16* (8), 5834-5847. DOI: https://doi.org/10.1016/j.rser.2012.06.004.

(2) Hermle, M.; Feldmann, F.; Bivour, M.; Goldschmidt, J. C.; Glunz, S. W. Passivating contacts and tandem concepts: Approaches for the highest silicon-based solar cell efficiencies. *Applied Physics Reviews* **2020**, *7* (2). DOI: 10.1063/1.5139202 (acccessed 1/2/2024).

(3) Shockley, W.; Queisser, H. J. Detailed Balance Limit of Efficiency of p-n Junction Solar Cells. *Journal of Applied Physics* **1961**, *32* (3), 510-519.

(4) Green, M. A.; Dunlop, E. D.; Hohl-Ebinger, J.; Yoshita, M.; Kopidakis, N.; Bothe, K.; Hinken, D.; Rauer, M.; Hao, X. Solar cell efficiency tables (Version 60). *Progress in Photovoltaics: Research and Applications* **2022**, *30* (7), 687-701.

(5) Sarkın, A. S.; Ekren, N.; Sağlam, Ş. A review of anti-reflection and self-cleaning coatings on photovoltaic panels. *Solar Energy* **2020**, *199*, 63-73.

(6) De Wolf, S.; Descoeudres, A.; Holman, Z. C.; Ballif, C. High-efficiency silicon heterojunction solar cells: A review. *green* **2012**, *2* (1), 7-24.

(7) Jariwala, D.; Sangwan, V. K.; Lauhon, L. J.; Marks, T. J.; Hersam, M. C. Carbon nanomaterials for electronics, optoelectronics, photovoltaics, and sensing. *Chemical Society Reviews* **2013**, *42* (7), 2824-2860.

(8) Yamaguchi, M.; Lee, K.-H.; Araki, K.; Kojima, N. A review of recent progress in heterogeneous silicon tandem solar cells. *Journal of Physics D: Applied Physics* **2018**, *51* (13), 133002.

(9) Hirst, L. C.; Ekins-Daukes, N. J. Fundamental losses in solar cells. *Progress in Photovoltaics: Research and Applications* **2011**, *19* (3), 286-293.

(10) Landsberg, P. T.; Badescu, V. Solar energy conversion: list of efficiencies and some theoretical considerations Part I—Theoretical considerations. *Progress in quantum electronics* **1998**, *22* (4), 211-230.

(11) Chin, X. Y.; Turkay, D.; Steele, J. A.; Tabean, S.; Eswara, S.; Mensi, M.; Fiala, P.; Wolff, C. M.; Paracchino, A.; Artuk, K. Interface passivation for 31.25%-efficient perovskite/silicon tandem solar cells. *Science* **2023**, *381* (6653), 59-63.

(12) Liu, C.; Yang, Y.; Chen, H.; Xu, J.; Liu, A.; Bati, A. S. R.; Zhu, H.; Grater, L.; Hadke, S. S.; Huang, C.; et al. Bimolecularly passivated interface enables efficient and stable inverted perovskite solar cells. *Science* **2023**, *382* (6672), 810-815. DOI: doi:10.1126/science.adk1633.

(13) Kim, K.-H.; Andreev, M.; Choi, S.; Shim, J.; Ahn, H.; Lynch, J.; Lee, T.; Lee, J.; Nazif, K. N.; Kumar, A. High-efficiency WSe2 photovoltaic devices with electron-selective contacts. *ACS nano* **2022**, *16* (6), 8827-8836.

(14) Jariwala, D.; Sangwan, V. K.; Lauhon, L. J.; Marks, T. J.; Hersam, M. C. Emerging device applications for semiconducting two-dimensional transition metal dichalcogenides. *ACS nano* **2014**, *8* (2), 1102-1120.

(15) Alfieri, A. D.; Ruth, T.; Lim, C.; Lynch, J.; Jariwala, D. Self-Hybridized Exciton-Polariton Photovoltaics. *arXiv preprint arXiv:2406.13065* **2024**.

(16) Amani, M.; Lien, D.-H.; Kiriya, D.; Xiao, J.; Azcatl, A.; Noh, J.; Madhvapathy, S. R.; Addou, R.; KC, S.; Dubey, M.; et al. Near-unity photoluminescence quantum yield in

MoS$_2$. *Science* **2015**, *350* (6264), 1065-1068. DOI: doi:10.1126/science.aad2114.
(17) Wang, L.; Huang, L.; Tan, W. C.; Feng, X.; Chen, L.; Huang, X.; Ang, K. W. 2D photovoltaic devices: progress and prospects. *Small Methods* **2018**, *2* (3), 1700294.
(18) Yu, Z. J.; Leilaeioun, M.; Holman, Z. Selecting tandem partners for silicon solar cells. *Nature Energy* **2016**, *1* (11), 1-4.
(19) Hu, Z.; Lin, D.; Lynch, J.; Xu, K.; Jariwala, D. How good can 2D excitonic solar cells be? *Device* **2023**.
(20) Wang, R.; Mujahid, M.; Duan, Y.; Wang, Z. K.; Xue, J.; Yang, Y. A review of perovskites solar cell stability. *Advanced Functional Materials* **2019**, *29* (47), 1808843.
(21) Laturia, A.; Van de Put, M. L.; Vandenberghe, W. G. Dielectric properties of hexagonal boron nitride and transition metal dichalcogenides: from monolayer to bulk. *npj 2D Materials and Applications* **2018**, *2* (1), 6.
(22) Kumar, P.; Lynch, J.; Song, B.; Ling, H.; Barrera, F.; Kisslinger, K.; Zhang, H.; Anantharaman, S. B.; Digani, J.; Zhu, H. Light–matter coupling in large-area van der Waals superlattices. *Nature nanotechnology* **2022**, *17* (2), 182-189.
(23) Pettersson, L. A.; Roman, L. S.; Inganäs, O. Modeling photocurrent action spectra of photovoltaic devices based on organic thin films. *Journal of Applied Physics* **1999**, *86* (1), 487-496.
(24) Hsu, C.; Frisenda, R.; Schmidt, R.; Arora, A.; De Vasconcellos, S. M.; Bratschitsch, R.; van der Zant, H. S.; Castellanos-Gomez, A. Thickness-dependent refractive index of 1L, 2L, and 3L MoS2, MoSe2, WS2, and WSe2. *Advanced optical materials* **2019**, *7* (13), 1900239.
(25) McPeak, K. M.; Jayanti, S. V.; Kress, S. J.; Meyer, S.; Iotti, S.; Rossinelli, A.; Norris, D. J. Plasmonic films can easily be better: rules and recipes. *ACS photonics* **2015**, *2* (3), 326-333.
(26) Grudinin, D.; Ermolaev, G.; Baranov, D.; Toksumakov, A.; Voronin, K.; Slavich, A.; Vyshnevyy, A.; Mazitov, A.; Kruglov, I.; Ghazaryan, D. Hexagonal boron nitride nanophotonics: a record-breaking material for the ultraviolet and visible spectral ranges. *Materials Horizons* **2023**.
(27) NREL. Reference Air Mass 1.5 Spectra. *Grid Mod.* **2019**, 4.
(28) Huang, R.; Yu, M.; Yang, Q.; Zhang, L.; Wu, Y.; Cheng, Q. Numerical simulation for optimization of an ultra-thin n-type WS2/p-type c-Si heterojunction solar cells. *Computational Materials Science* **2020**, *178*, 109600.
(29) Limpert, S.; Ghosh, K.; Wagner, H.; Bowden, S.; Honsberg, C.; Goodnick, S.; Bremner, S.; Ho-Baillie, A.; Green, M. Results from coupled optical and electrical sentaurus TCAD models of a gallium phosphide on silicon electron carrier selective contact solar cell. In *2014 IEEE 40th Photovoltaic Specialist Conference (PVSC)*, 2014; IEEE: pp 0836-0840.
(30) Tian, H.; Chin, M. L.; Najmaei, S.; Guo, Q.; Xia, F.; Wang, H.; Dubey, M. Optoelectronic devices based on two-dimensional transition metal dichalcogenides. *Nano Research* **2016**, *9*, 1543-1560.
(31) Adachi, S. *Properties of semiconductor alloys: group-IV, III-V and II-VI semiconductors*; John Wiley & Sons, 2009.
(32) Baines, T.; Shalvey, T. P.; Major, J. D. Cdte solar cells. In *A comprehensive guide to solar energy systems*, Elsevier, 2018; pp 215-232.
(33) Marti, A.; Araújo, G. L. Limiting efficiencies for photovoltaic energy conversion in

multigap systems. *Solar Energy Materials and Solar Cells* **1996**, *43* (2), 203-222.

(34) Aeberhard, U.; Altazin, S.; Stepanova, L.; Stous, A.; Blülle, B.; Kirsch, C.; Knapp, E.; Ruhstaller, B. Numerical Optimization of Organic and Hybrid Multijunction Solar Cells. In *2019 IEEE 46th Photovoltaic Specialists Conference (PVSC)*, 2019; IEEE: pp 0105-0111.

(35) Raja, A.; Chaves, A.; Yu, J.; Arefe, G.; Hill, H. M.; Rigosi, A. F.; Berkelbach, T. C.; Nagler, P.; Schüller, C.; Korn, T. Coulomb engineering of the bandgap and excitons in two-dimensional materials. *Nature communications* **2017**, *8* (1), 15251.

(36) Shi, E.; Zhang, L.; Li, Z.; Li, P.; Shang, Y.; Jia, Y.; Wei, J.; Wang, K.; Zhu, H.; Wu, D. TiO2-coated carbon nanotube-silicon solar cells with efficiency of 15%. *Scientific reports* **2012**, *2* (1), 884.

(37) Atwater, H. A.; Polman, A. Plasmonics for improved photovoltaic devices. *Nature Materials* **2010**, *9* (3), 205-213. DOI: 10.1038/nmat2629.

(38) Nassiri Nazif, K.; Nitta, F. U.; Daus, A.; Saraswat, K. C.; Pop, E. Efficiency limit of transition metal dichalcogenide solar cells. *Communications Physics* **2023**, *6* (1), 367. DOI: 10.1038/s42005-023-01447-y.

(39) Liu, Y.; Guo, J.; Zhu, E.; Liao, L.; Lee, S.-J.; Ding, M.; Shakir, I.; Gambin, V.; Huang, Y.; Duan, X. Approaching the Schottky–Mott limit in van der Waals metal–semiconductor junctions. *Nature* **2018**, *557* (7707), 696-700. DOI: 10.1038/s41586-018-0129-8.

(40) Svatek, S. A.; Bueno-Blanco, C.; Lin, D.-Y.; Kerfoot, J.; Macías, C.; Zehender, M. H.; Tobías, I.; García-Linares, P.; Taniguchi, T.; Watanabe, K.; et al. High open-circuit voltage in transition metal dichalcogenide solar cells. *Nano Energy* **2021**, *79*, 105427. DOI: https://doi.org/10.1016/j.nanoen.2020.105427.

(41) Hanbicki, A.; Currie, M.; Kioseoglou, G.; Friedman, A.; Jonker, B. Measurement of high exciton binding energy in the monolayer transition-metal dichalcogenides WS2 and WSe2. *Solid State Communications* **2015**, *203*, 16-20.

(42) Shen, C.-C.; Hsu, Y.-T.; Li, L.-J.; Liu, H.-L. Charge dynamics and electronic structures of monolayer MoS2 films grown by chemical vapor deposition. *Applied Physics Express* **2013**, *6* (12), 125801.

(43) Ugeda, M. M.; Bradley, A. J.; Shi, S.-F.; da Jornada, F. H.; Zhang, Y.; Qiu, D. Y.; Ruan, W.; Mo, S.-K.; Hussain, Z.; Shen, Z.-X.; et al. Giant bandgap renormalization and excitonic effects in a monolayer transition metal dichalcogenide semiconductor. *Nature Materials* **2014**, *13* (12), 1091-1095. DOI: 10.1038/nmat4061.

(44) Essig, S.; Allebé, C.; Remo, T.; Geisz, J. F.; Steiner, M. A.; Horowitz, K.; Barraud, L.; Ward, J. S.; Schnabel, M.; Descoeudres, A.; et al. Raising the one-sun conversion efficiency of III–V/Si solar cells to 32.8% for two junctions and 35.9% for three junctions. *Nature Energy* **2017**, *2* (9), 17144. DOI: 10.1038/nenergy.2017.144.

(45) Mariotti, S.; Köhnen, E.; Scheler, F.; Sveinbjörnsson, K.; Zimmermann, L.; Piot, M.; Yang, F.; Li, B.; Warby, J.; Musiienko, A.; et al. Interface engineering for high-performance, triple-halide perovskite–silicon tandem solar cells. *Science* **2023**, *381* (6653), 63-69. DOI: doi:10.1126/science.adf5872.

(46) Liu, J.; Shi, B.; Xu, Q.; Li, Y.; Li, Y.; Liu, P.; SunLi, Z.; Wang, X.; Sun, C.; Han, W.; et al. Textured Perovskite/Silicon Tandem Solar Cells Achieving Over 30% Efficiency Promoted by 4-Fluorobenzylamine Hydroiodide. *Nano-Micro Letters* **2024**, *16* (1), 189. DOI: 10.1007/s40820-024-01406-4.

(47) Tamboli, A. C.; Bobela, D. C.; Kanevce, A.; Remo, T.; Alberi, K.; Woodhouse, M. Low-Cost CdTe/Silicon Tandem Solar Cells. *IEEE Journal of Photovoltaics* **2017**, *7* (6), 1767-

1772. DOI: 10.1109/JPHOTOV.2017.2737361.
(48) Heriche, H.; Rouabah, Z.; Bouarissa, N. New ultra thin CIGS structure solar cells using SCAPS simulation program. *International Journal of Hydrogen Energy* **2017**, *42* (15), 9524-9532. DOI: https://doi.org/10.1016/j.ijhydene.2017.02.099.

Supporting Information

Tandem Photovoltaics from 2D Transition Metal Dichalcogenides on Silicon

Zekun Hu[1], Sudong Wang[1], Jason Lynch[1], Deep Jariwala[1,*]

[1]Department of Electrical and Systems Engineering, University of Pennsylvania, Philadelphia, PA, USA

Contents

Number of pages: 12

Number of figures: 6

Number of tables: 4

1. Top Cell Electrical Simulation Model

The two-dimensional p-i-n superlattice structure with separated contacts was simulated using Sentaurus technology computer aided design tools from Synopsys. Sentaurus offers a comprehensive understanding of device behavior by incorporating a multitude of physical mechanisms, including free carrier recombination processes and exciton dissociation process. The simulations were conducted with a default depth of 1 μm in the third dimension, with no significant impact on the outcomes expected from variations in this dimension.

The top structure consists of the monolayer active materials ($MoS_2$, $MoSe_2$, $WS_2$, $WSe_2$), insulators ($Al_2O_3$), cathode (Ag), and anode (Au). The effects of the thick bottom insulator were omitted from the Sentaurus simulation, as these were adequately addressed within the photon generation simulation parameters. The model accounted for a variety of factors influencing device performance, including binding energy, exciton diffusion length, exciton lifetimes (radiative and nonradiative), free carrier mobility, free carrier Shockley-Read-Hall (SRH) lifetime, and device length.

Initial parameters were derived from empirical data and theoretical calculations to define the materials and the overall device structure. Electron and hole densities were ascertained using quasi-Fermi potentials, and bandgap values were adopted from established literature. The model accommodated discontinuities at the superlattice interfaces, enabling the accurate representation of heterointerfaces through double points in the dataset.

Optical generation rates, calculated by transfer matrix method, were assigned for each layer, and carrier recombination was modeled to include SRH, Auger, and radiative processes. The model's coupled equations—Poisson, electron and hole continuity, and the singlet exciton equation—described the exciton dynamics within the device:

$$\frac{\partial n_{\text{se}}}{\partial t} = R_{\text{bimolec}} + \nabla * D_{\text{se}} \nabla n_{\text{se}} - \frac{n_{\text{se}} - n_{\text{se}}^{\text{eq}}}{\tau} - \frac{n_{\text{se}} - n_{\text{se}}^{\text{eq}}}{\tau_{\text{trap}}} - R_{\text{se}} \qquad \text{Eq.1}$$

Where $n_{se}$ and $n_{\text{se}}^{\text{eq}}$ are the singlet exciton and equilibrium exciton densities, $R_{\text{bimolec}}$ is the carrier bimolecular recombination rate acting as a singlet exciton generation term, $D_{\text{se}}$ is the singlet exciton diffusion constant, $\tau$, $\tau_{\text{trap}}$ are the singlet exciton lifetimes. $R_{\text{se}}$ is the net singlet exciton recombination rate.

The Langevin recombination model was applied for electron-hole pair and exciton recombination, with the rate:

$$R_{\text{bimolec}} = \gamma * \frac{q}{\varepsilon_0 \varepsilon_r} * (\mu_n + \mu_p)\left(np - n_{\text{i,eff}}^2 \frac{n_{\text{se}}}{n_{\text{se}}^{\text{eq}}}\right) \qquad \text{Eq.2}$$

Where $\gamma$ is 0.25, $q$ is the elementary charge, $\varepsilon_0$ and $\varepsilon_r$ denote the free space and relative permittivity, respectively. Electron and hole mobilities are given by $\mu_n$ and $\mu_p$, accordingly. $n$, $p$, and $n_{\text{i,eff}}$ describe the electron, hole, and effective intrinsic density, respectively. $n_{\text{se}}$ is the singlet exciton density and $n_{se}^{eq}$ denotes the singlet-exciton equilibrium density.

The boundary and continuity conditions for singlet exciton equation at electrodes were specified by:

$$n_{se}(T) = n_{se}^{eq}(T) = \gamma g_{ex}(N_C(T) + N_V(T))\exp\left(-\frac{E_{g,ele} - E_{ex}}{kT}\right) \qquad \text{Eq.3}$$

Where $g_{ex}$ and $E_{ex}$ are the singlet exciton degeneracy factor and binding energy, respectively. $N_C(T)$ and $N_V(T)$ are the electron and hole effective density-of-states, accordingly. $E_{g,ele}$ is the electric band gap, k is the Boltzmann constant and T is the temperature.

The exciton dissociation model describing the dissociation of singlet excitons into electron-hole pairs at semiconductor-semiconductor and semiconductor-insulator interfaces is given by:

$$R_{se,diss}^{surf} = v_{se}\sigma_{se-N_{diss}}N_{se,diss}^{surf}(n_{se} - n_{se}^{eq}) \qquad \text{Eq.4}$$

Where $R_{se,diss}^{surf}$ is the rate of singlet exciton interface dissociation, $v_{se}$ is the velocity of singlet exciton, $\sigma_{se-N_{diss}}$ is the capture cross-section of exciton dissociation centers with the surface density $N_{se,diss}^{surf}$.

The relation between electrical band gap and optical band gap is:

$$E_{g,ele} = E_{g,opt} + E_b \qquad \text{Eq.5}$$

Where $E_{g,opt}$ is the optical band gap, $E_b$ is the exciton binding energy.

2. Bottom cell simulation model

The bottom section of the tandem solar cell was modeled as a monocrystalline silicon (mono-Si) solar cell with a rear-contact configuration. The simulated structure encompassed the top and bottom contacts, a surface nitride layer, and the active silicon layer. Mesh optimization was particularly focused on the area beneath the top contact and in the vicinity of the rear contacts to ensure precise simulation results.

In contrast to the top cell simulation, where excitonic effects are crucial, the bottom Si cell was modeled using a standard drift-diffusion model(1), without the need to account for excitonic behavior. To realize a single-junction silicon solar cell with a PCE of 23.28%, the design parameters underwent extensive optimization. The cell featured a 150-micron-thick silicon layer, a 10-micron-long top contact, and a rear contact offset from the front contact by 260 microns. The total area of the rear contact was set to 50 µm². Recombination dynamics were captured by considering both SRH and Auger lifetimes, with the surface recombination rate set to $10^5$ cm/s and the contact resistivity to $1\times10^{-6}$ Ω·cm². The bulk lifetime of silicon was fixed at 200 µs.

When the active Auger recombination is considered, the single-junction silicon solar cell exhibited a decrease in PCE by 0.34%, dropping to 22.94%. This highlights the impact of recombination mechanisms on the overall efficiency of silicon-based solar cells and underscores the necessity of careful lifetime management to achieve high-performance photovoltaic devices.

3. Supporting Figures

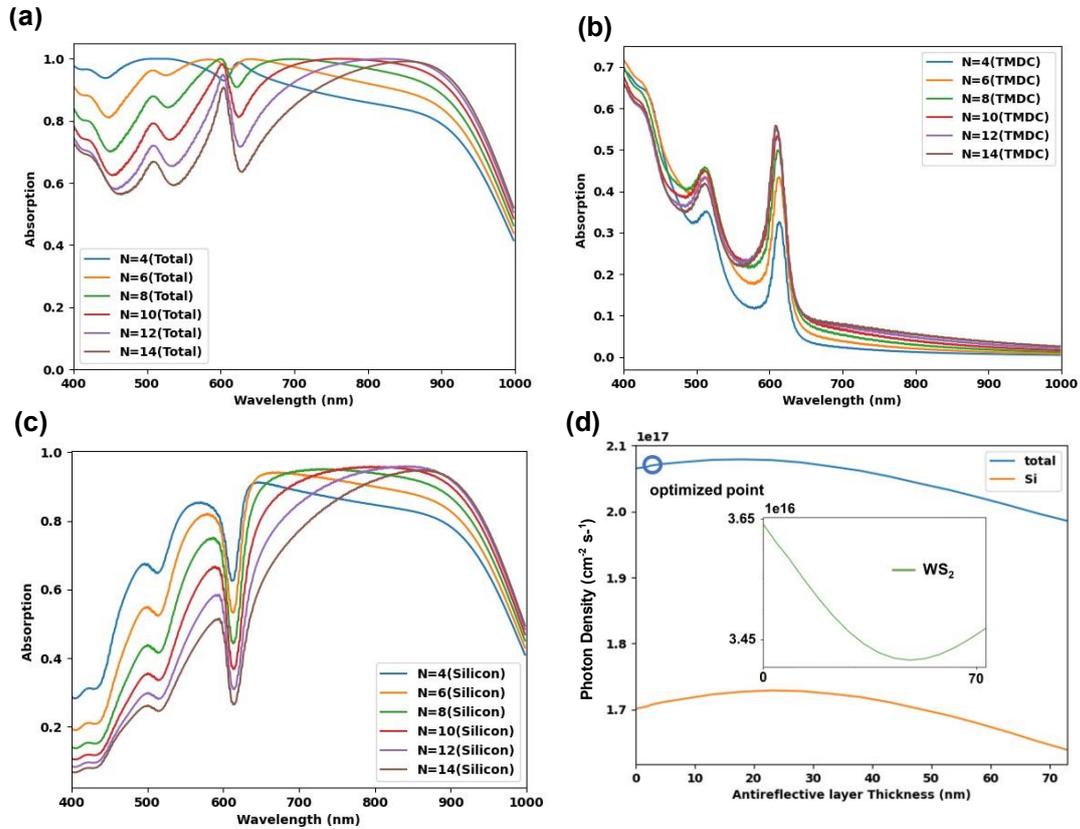

Figure S1. (a) Optimization of the superlattice layer number based on $WS_2$ without optimizing bottom insulator thickness. (b) Detailed absorption spectrum in $WS_2$ top cell part. (c) Detailed absorption spectrum in Si bottom cell part. (d) Optimization of the antireflective layer's thickness above the superlattice to augment absorption based on $WS_2$, with an inset displaying a detailed absorption of the top cell region.

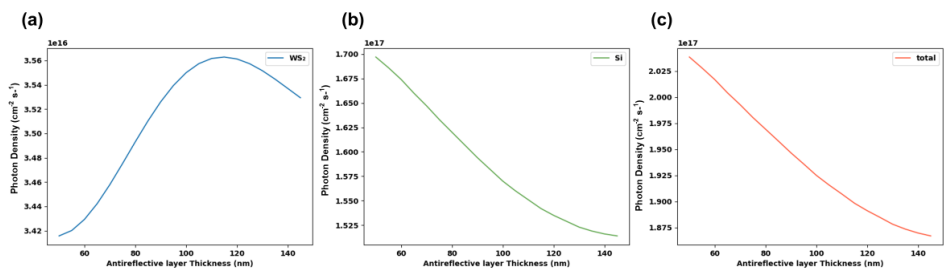

Figure S2. Optimization of the antireflective layer's thickness above the superlattice to augment absorption in extended range based on $WS_2$, (a) top cell in blue, (b) bottom cell in green, (c) total cell in orange.

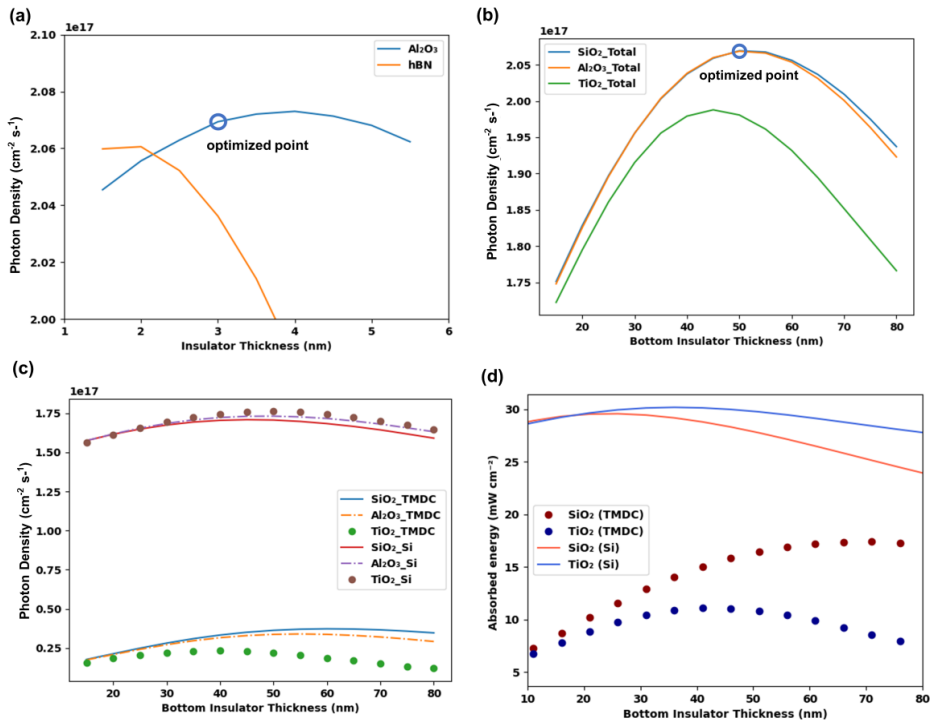

Figure S3. (a) Comparative analysis of hBN and Al$_2$O$_3$ as insulators between WS$_2$ monolayers to maximize total photon absorption. (b) Evaluation of SiO$_2$, Al$_2$O$_3$, and TiO$_2$ insulators for optimal interfacing between the TMDC superlattice and the Si solar cell based on WS$_2$. (c) Detailed photon absorption in the top and bottom cells influenced by various insulator materials based on WS$_2$. (d) Detailed photon absorption in the top and bottom cells influenced by various insulator materials based on MoSe$_2$.

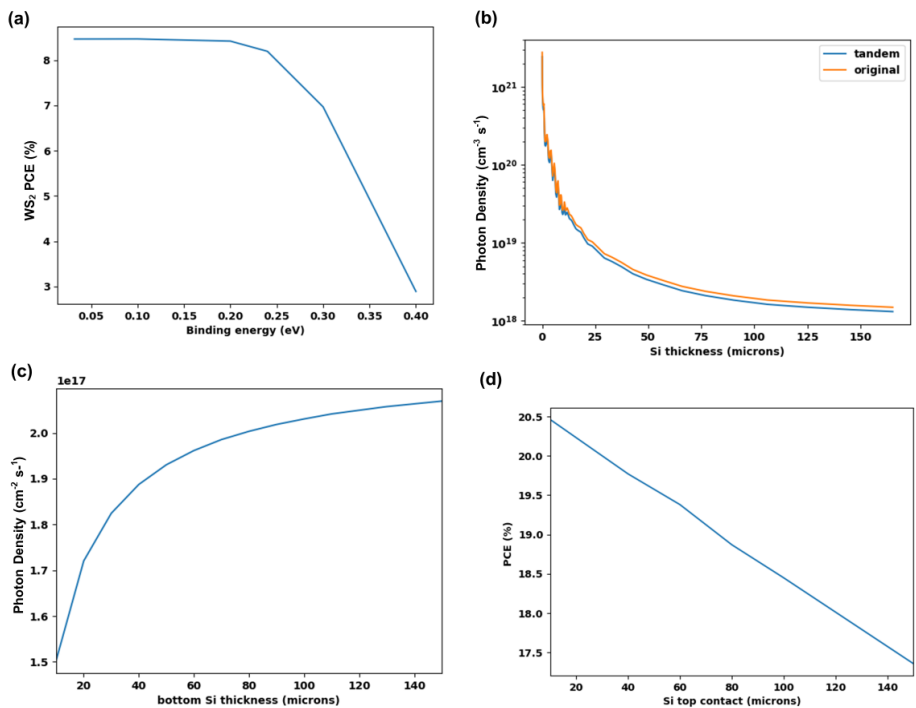

Figure S4. (a) Top cell power conversion efficiency with varying binding energy of WS$_2$. The free carrier mobility is assumed as 1000 cm$^2$ v$^{-1}$ s$^{-1}$. (b) Optimization of photon absorption in different incident angles. (c) Optimization of silicon photon absorption by varying Silicon thickness. (d) Optimization of bottom cell efficiency in terms of top contact length (total cell length is 500 microns).

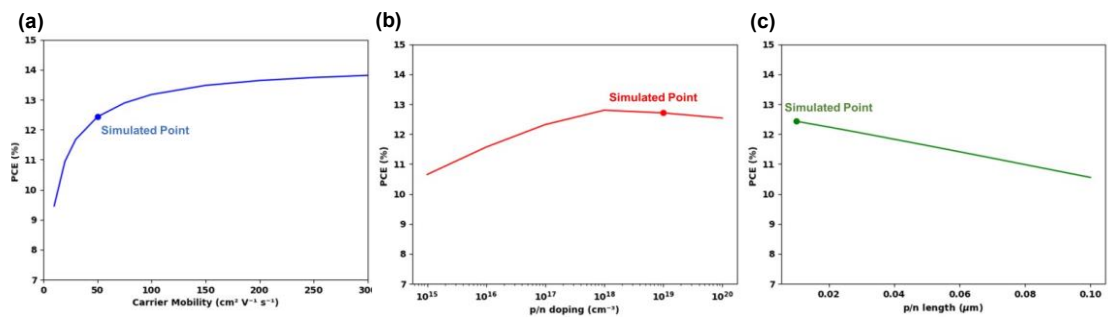

Figure S5. (a) Optimization of top cell efficiency of MoSe$_2$ in terms of carrier mobility. (b) Optimization of top cell efficiency of MoSe$_2$ in terms of p-/n- doping concentration. (c) Optimization of top cell efficiency of MoSe$_2$ in terms of p-/n- region length (total cell length is 1 micron). Parameters used for the main results are labelled as 'Simulated Point'.

4. Top Cell Material parameters

Table S1. Parameters of monolayer TMDCs.

| Material (monolayer) | $MoS_2$ | $MoSe_2$ | $WS_2$ | $WSe_2$ | $MoTe_2$ |
|---|---|---|---|---|---|
| Optical Bandgap (eV) | 1.8(2) | 1.55(3) | 2.04(4) | 1.65(5) | 1.19(6) |
| Binding energy (eV) | 0.48(7) | 0.57(8,9) | 0.32(10) | 0.37(10) | - |
| Exciton Diffusion length (um) | 1.5(11) | 1.22(12) | 0.35(13) | 0.16(14) | - |
| Radiative exciton lifetime (ns) | 8(11) | 0.8(15) | 4.4(15) | 3.5(15) | - |
| Free carrier lifetime (ns) | 10(11) | 130(12) | 22(13) | 18(14) | - |
| Free carrier mobility ($cm^2$/Vs) | 60(16) | 50(17) | 33(18) | 82(19) | - |

Table S2. Geometry and simulation parameters for the top cell.

| Default Model of simulation | |
|---|---|
| Accepter Concentration ($cm^{-3}$) | $10^{19}$ |
| Donor Concentration ($cm^{-3}$) | $10^{19}$ |
| Temperature (K) | 300 |
| Electron lifetime (ns) | 1.5 |
| Hole Lifetime (ns) | 1.5 |
| Active region length (um) | 1 |

Table S3. Parameters of bulk materials.

| Material (bulk) | MoSe$_2$ |
|---|---|
| Bandgap (eV) | 1.1(3) |
| Binding energy (eV) | 0.1(20) |
| Exciton Diffusion length (um) | 0.6(12) |
| Radiative exciton lifetime (ns) | 0.21(12) |
| Free carrier lifetime (ns) | 1.5 |
| Free carrier mobility (cm$^2$/Vs) | 10 |

5. Bottom Cell parameters

Table S4. Parameters of monolayer TMDCs.

| Material | Silicon |
|---|---|
| Bandgap (eV) | 1.12 |
| Bulk lifetime (us) | 200 |
| SRH recombination rate (cm/s) | $10^5$ |

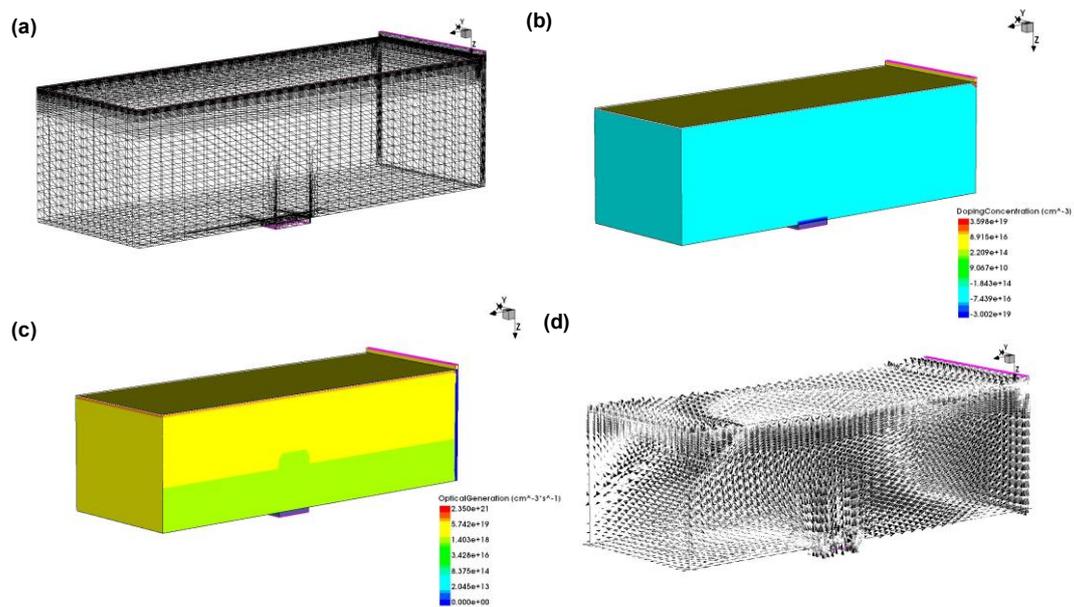

Figure S6. Simulation details of bottom cell. (a) Meshing refinement near contacts and top surface. (b) Doping profile. (c) Optical generation rate with consideration of different bottom reflectivity. (d) Total current vector shown in black solid arrows.

6. Reference:

(1) Benaichi M.; Chetouani A.; Karkri A.; Moussaid D.; Elqabbaj S. E. Three-Dimensional Drift-Diffusion Model for Simulation and Investigation of Bordering Effects in Silicon Solar Cells. *Materials Today: Proceedings* **2019**, *13*, 630-636. DOI: https://doi.org/10.1016/j.matpr.2019.04.022.

(2) Castellanos-Gomez A.; Barkelid M.; Goossens A.; Calado V. E.; van der Zant H. S.; Steele G. A. Laser-thinning of MoS2: on demand generation of a single-layer semiconductor. *Nano letters* **2012**, *12* (6), 3187-3192.

(3) Tongay S.; Zhou J.; Ataca C.; Lo K.; Matthews T. S.; Li J.; Grossman J. C.; Wu J. Thermally driven crossover from indirect toward direct bandgap in 2D semiconductors: MoSe2 versus MoS2. *Nano letters* **2012**, *12* (11), 5576-5580.

(4) Cheng G.; Li B.; Zhao C.; Jin Z.; Li H.; Lau K. M.; Wang J. Exciton aggregation induced photoluminescence enhancement of monolayer WS2. *Applied Physics Letters* **2019**, *114* (23).

(5) He K.; Kumar N.; Zhao L.; Wang Z.; Mak K. F.; Zhao H.; Shan J. Tightly bound excitons in monolayer WSe 2. *Physical review letters* **2014**, *113* (2), 026803.

(6) Li J.-H.; Bing D.; Wu Z.-T.; Wu G.-Q.; Bai J.; Du R.-X.; Qi Z.-Q. Thickness-dependent excitonic properties of atomically thin 2H-MoTe2. *Chinese Physics B* **2020**, *29* (1), 017802.

(7) Shen C.-C.; Hsu Y.-T.; Li L.-J.; Liu H.-L. Charge dynamics and electronic structures of monolayer MoS2 films grown by chemical vapor deposition. *Applied Physics Express* **2013**, *6* (12), 125801.

(8) Ghosh C.; Sarkar D.; Mitra M.; Chattopadhyay K. Equibiaxial strain: tunable electronic structure and optical properties of bulk and monolayer MoSe2. *Journal of Physics D: Applied Physics* **2013**, *46* (39), 395304.

(9) Ugeda M. M.; Bradley A. J.; Shi S.-F.; da Jornada F. H.; Zhang Y.; Qiu D. Y.; Ruan W.; Mo S.-K.; Hussain Z.; Shen Z.-X.; et al. Giant bandgap renormalization and excitonic effects in a monolayer transition metal dichalcogenide semiconductor. *Nature Materials* **2014**, *13* (12), 1091-1095. DOI: 10.1038/nmat4061.

(10) Hanbicki A.; Currie M.; Kioseoglou G.; Friedman A.; Jonker B. Measurement of high exciton binding energy in the monolayer transition-metal dichalcogenides WS2 and WSe2. *Solid State Communications* **2015**, *203*, 16-20.

(11) Uddin S. Z.; Kim H.; Lorenzon M.; Yeh M.; Lien D.-H.; Barnard E. S.; Htoon H.; Weber-Bargioni A.; Javey A. Neutral exciton diffusion in monolayer MoS2. *ACS nano* **2020**, *14* (10), 13433-13440.

(12) Kumar N.; Cui Q.; Ceballos F.; He D.; Wang Y.; Zhao H. Exciton diffusion in monolayer and bulk MoSe 2. *Nanoscale* **2014**, *6* (9), 4915-4919.

(13) He J.; He D.; Wang Y.; Cui Q.; Ceballos F.; Zhao H. Spatiotemporal dynamics of excitons in monolayer and bulk WS2. *Nanoscale* **2015**, *7* (21), 9526-9531.

(14) Cui Q.; Ceballos F.; Kumar N.; Zhao H. Transient absorption microscopy of monolayer and bulk WSe2. *ACS nano* **2014**, *8* (3), 2970-2976.

(15) Palummo M.; Bernardi M.; Grossman J. C. Exciton radiative lifetimes in two-dimensional transition metal dichalcogenides. *Nano letters* **2015**, *15* (5), 2794-2800.

(16) Huo N.; Yang Y.; Wu Y.-N.; Zhang X.-G.; Pantelides S. T.; Konstantatos G. High carrier mobility in monolayer CVD-grown MoS 2 through phonon suppression. *Nanoscale* **2018**, *10* (31), 15071-15077.

(17) Wang X.; Gong Y.; Shi G.; Chow W. L.; Keyshar K.; Ye G.; Vajtai R.; Lou J.; Liu Z.; Ringe E.; et al. Chemical Vapor Deposition Growth of Crystalline Monolayer MoSe2. *ACS Nano* **2014**, *8* (5), 5125-5131. DOI: 10.1021/nn501175k.

(18) Sebastian A.; Pendurthi R.; Choudhury T. H.; Redwing J. M.; Das S. Benchmarking monolayer MoS2 and WS2 field-effect transistors. *Nature Communications* **2021**, *12* (1), 693. DOI: 10.1038/s41467-020-20732-w.

(19) Ji H. G.; Solís-Fernández P.; Yoshimura D.; Maruyama M.; Endo T.; Miyata Y.; Okada S.; Ago H. Chemically Tuned p- and n-Type WSe2 Monolayers with High Carrier Mobility for Advanced Electronics. *Advanced Materials* **2019**, *31* (42), 1903613. DOI: https://doi.org/10.1002/adma.201903613.

(20) Arora A.; Nogajewski K.; Molas M.; Koperski M.; Potemski M. Exciton band structure in layered MoSe 2: from a monolayer to the bulk limit. *Nanoscale* **2015**, *7* (48), 20769-20775.